# Iterative Conditional Fitting for Gaussian Ancestral Graph Models

**Mathias Drton**
Department of Statistics
University of Washington
Seattle, WA 98195-4322

**Thomas S. Richardson**
Department of Statistics
University of Washington
Seattle, WA 98195-4322

## Abstract

Ancestral graph models, introduced by Richardson and Spirtes (2002), generalize both Markov random fields and Bayesian networks to a class of graphs with a global Markov property that is closed under conditioning and marginalization. By design, ancestral graphs encode precisely the conditional independence structures that can arise from Bayesian networks with selection and unobserved (hidden/latent) variables. Thus, ancestral graph models provide a potentially very useful framework for exploratory model selection when unobserved variables might be involved in the data-generating process but no particular hidden structure can be specified. In this paper, we present the Iterative Conditional Fitting (ICF) algorithm for maximum likelihood estimation in Gaussian ancestral graph models. The name reflects that in each step of the procedure a conditional distribution is estimated, subject to constraints, while a marginal distribution is held fixed. This approach is in duality to the well-known Iterative Proportional Fitting algorithm, in which marginal distributions are fitted while conditional distributions are held fixed.

## 1 INTRODUCTION

Markov random fields or equivalently undirected graph models as well as Bayesian networks or equivalently directed acyclic directed graph (DAG) models have found wide-spread application. Well-known generalizations of both undirected graph models and DAG models are the chain graph models, which can be equipped with two alternative Markov properties (Andersson et al. 2001). A different generalization is obtained from ancestral graphs, introduced by Richardson and Spirtes (2002) = RS (2002). Whereas chain graphs allow both undirected and directed edges, ancestral graphs have edges of three possible types: undirected and directed edges are complemented by bidirected edges.

In ancestral graphs, $m$-separation, a natural extension of $d$-separation, yields a global Markov property that is closed under conditioning and marginalization. Interpreted via this Markov property, ancestral graphs encode precisely the conditional independence structures that can arise from a Bayesian network with selection and unobserved variables (RS 2002). Marginalization (forming the marginal distribution of the observed variables) is associated with introducing bi-directed edges; conditioning (on selection variables) is associated with introducing undirected edges. Due to their connection with underlying DAGs, ancestral graphs encode causally interpretable conditional independence structures. This feature is attractive for exploratory model selection when the presence of unobserved variables cannot be excluded but no detailed knowledge about the structure of these unobserved variables is available.

To illustrate the conditioning and marginalization connection between DAGs and ancestral graphs, consider the DAG in Figure 1. Assume that variables $u_{23}$ and $u_{34}$ are unobserved and that variable $s_{01}$ is a selection variable. If we form the conditional distribution $(0\ 1\ 2\ 3\ 4 \mid s_{01})$, with unobserved variables marginalized out and selection variables conditioned on, then the conditional independences holding in $(0\ 1\ 2\ 3\ 4 \mid s_{01})$ are exactly those encoded by the ancestral graph $G$ from Figure 2. RS (2002) give details on this connection between DAGs and ancestral graphs.

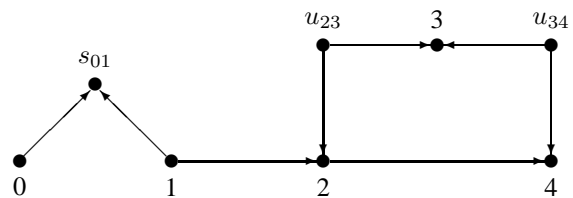

Figure 1: DAG with selection variable $s_{01}$ and the unobserved variables $u_{23}$ and $u_{34}$.

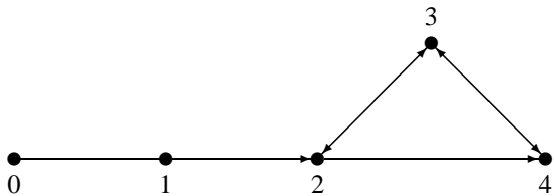

Figure 2: An ancestral graph.

This paper is a first step towards making ancestral graph methodology available for use in applications. We consider the problem of estimating (learning) the parameters of a given ancestral graph model by maximum likelihood. We restrict ourselves to the Gaussian case, for which RS (2002) provided a parameterization, and propose an algorithm, which extends previous work by Drton and Richardson (2003). We name this algorithm Iterative Conditional Fitting (ICF) since in each step of the procedure a conditional distribution is estimated, subject to constraints, while a marginal distribution is held fixed. This approach is in duality to the well-known iterative proportional fitting algorithm (Whittaker 1990, pp. 182–185), in the steps of which a marginal distribution is fitted for a fixed conditional distribution.

This paper is organized as follows. In §§2 and 3 we define ancestral graphs and their global Markov property, and in §4 we introduce Gaussian ancestral graph models. In §5 we present the ICF algorithm. An implementation in the R statistical programming system is demonstrated in §6. Experiments are described in §7. We conclude in §8.

## 2 ANCESTRAL GRAPHS

Consider a graph $G = (V, E)$ with the vertex set $V$ and the edge set $E$ containing three types of edge, *undirected* ($-$), *directed* ($\rightarrow$) and *bidirected* ($\leftrightarrow$). The graph $G$ is not allowed to have an edge from a vertex to itself or more than one edge between a given pair of vertices. The following terminology describes relations between two vertices $i, j \in V$ in $G$:

If $\left\{ \begin{array}{c} i - j \\ i \leftrightarrow j \\ i \rightarrow j \end{array} \right\}$ then $i$ is a $\left\{ \begin{array}{c} \text{neighbor} \\ \text{spouse} \\ \text{parent} \end{array} \right\}$ of $j$.

The set of neighbors, spouses and parents of a vertex $i$ are denoted by $\text{ne}(i)$, $\text{sp}(i)$ and $\text{pa}(i)$, respectively.

A sequence of edges between two vertices $i$ and $j$ in $G$ is an ordered (multi)set of edges $< e_1, \ldots, e_m >$, such that there exists a sequence of vertices (not necessarily distinct) $< i = i_1, \ldots, i_{m+1} = j >$, where edge $e_{i_k}$ has endpoints $i_k, i_{k+1}$. A sequence of edges for which the corresponding sequence of vertices contains no repetitions is called a *path*. A path of the form $i \rightarrow \cdots \rightarrow j$, on which every edge is of the form $\rightarrow$, with the arrowheads pointing toward $j$, is a *directed path from $i$ to $j$*. A vertex $i$ is said to be an *ancestor* of a vertex $j$, denoted $i \in \text{an}(j)$, if either there is a directed path $i \rightarrow \cdots \rightarrow j$ from $i$ to $j$, or $i = j$. For vertex sets $A \subseteq V$, we define $\text{an}(A) = \cup(\text{an}(i) \mid i \in A)$ and similarly $\text{sp}(A)$, $\text{pa}(A)$, $\text{ne}(A)$.

**Definition 1 (Ancestral Graphs, RS 2002, §3)** *A graph $G = (V, E)$ with undirected, directed and bidirected edges is an* ancestral graph *if for all $i \in V$ it holds that*

*(i) if* $\text{ne}(i) \neq \emptyset$ *then* $\text{pa}(i) \cup \text{sp}(i) = \emptyset$;

*(ii)* $i \notin \text{an}(\text{pa}(i) \cup \text{sp}(i))$.

Condition (i) states that if there exists an undirected edge with endpoint $i$ then there may not exist a directed or bidirected edge with an arrowhead at $i$, and condition (ii) states that there may not be a directed path from a vertex $i$ to one of its parents or spouses.

An example of an ancestral graph with vertex set $V = \{0, 1, 2, 3, 4\}$ is given in Figure 2. Additional examples can be found in RS (2002, e.g. Fig. 3, 6, 7 and 12). Note also that directed acyclic graphs (DAGs) and undirected graphs are ancestral graphs.

Condition (i) implies that an ancestral graph can be decomposed into an undirected part and a part with only directed and bidirected edges (RS 2002, §3.2). Let $\text{un}_G = \{i \in V \mid \text{pa}(i) \cup \text{sp}(i) = \emptyset\}$. Then the subgraph $G_{\text{un}_G} = [\text{un}_G, E \cap (\text{un}_G \times \text{un}_G)]$ induced by $\text{un}_G$ is an undirected graph, and any edge between $i \in \text{un}_G$ and $j \notin \text{un}_G$ is directed as $i \rightarrow j$. Furthermore, the induced subgraph $G_{V \setminus \text{un}_G}$ contains only directed and bidirected edges. In Figure 2, $\text{un}_G = \{0, 1\}$.

## 3 GLOBAL MARKOV PROPERTY

Pearl's (1988) $d$-separation criterion for DAGs can be extended to ancestral graphs. A nonendpoint vertex $i$ on a path is a *collider on the path* if the edges preceding and succeeding $i$ on the path have an arrowhead at $i$, that is, $\rightarrow i \leftarrow$, $\rightarrow i \leftrightarrow$, $\leftrightarrow i \leftarrow$, or $\leftrightarrow i \leftrightarrow$. A nonendpoint vertex $i$ on a path which is not a collider is a *noncollider on the path*. A path between vertices $i$ and $j$ in an ancestral graph $G$ is said to be *m-connecting given a set $C$* (possibly empty), with $i, j \notin C$, if:

(i) every noncollider on the path is not in $C$, and

(ii) every collider on the path is in $\text{an}(C)$.

If there is no path $m$-connecting $i$ and $j$ given $C$, then $i$ and $j$ are said to be $m$-separated given $C$. Sets $A$ and $B$ are $m$-separated given $C$, if for every pair $i, j$, with $i \in A$ and $j \in B$, $i$ and $j$ are $m$-separated given $C$ ($A, B, C$ are disjoint sets; $A, B$ are nonempty). This is an extension of Pearl's $d$-separation criterion to ancestral graphs in that in a DAG, a path is $d$-connecting if and only if it is $m$-connecting.

Let $G = (V, E)$ be an ancestral graph whose vertices $V$ are identified with random variables $(Y_i \mid i \in V)$ and let $P$ be the joint probability distribution of $(Y_i \mid i \in V)$. If $Y_A \perp\!\!\!\perp Y_B \mid Y_C$ whenever $A$ and $B$ are $m$-separated given $C$, then $P$ is said to satisfy the *global Markov property for* $G$, or to be *globally Markov with respect to* $G$. Here $Y_A = (Y_i \mid i \in A)$ for $A \subseteq V$. For the joint distribution $P$ to be globally Markov with respect to the graph $G$ in Figure 2, the conditional independences $Y_0 \perp\!\!\!\perp Y_{234} \mid Y_1$, $Y_{01} \perp\!\!\!\perp Y_3$, $Y_{01} \perp\!\!\!\perp Y_4 \mid Y_2$ must hold.

In this example, if two vertices $i$ and $j$ are not adjacent, then a conditional independence $Y_i \perp\!\!\!\perp Y_j \mid Y_C$, for some $C \subseteq V$, holds. Ancestral graphs for which this is true are called *maximal* (RS 2002, §3.7). In fact, by adding bidirected edges, any non-maximal ancestral graph can be converted into a unique maximal ancestral graph without changing the independence model implied by the global Markov property.

## 4  GAUSSIAN ANCESTRAL GRAPH MODELS

Suppose now that the variables $(Y_i \mid i \in V)$ jointly follow a centered Gaussian $\equiv$ normal distribution $\mathcal{N}_V(0, \Sigma) \in \mathbb{R}^V$, where $\Sigma = (\sigma_{ij}) \in \mathbb{R}^{V \times V}$ is the unknown positive definite covariance matrix. Let $\mathbf{P}(V)$ be the set of all positive definite $V \times V$ matrices and $\mathbf{P}(G)$ the subset of all $\Sigma \in \mathbf{P}(V)$ such that $\mathcal{N}_V(0, \Sigma)$ is globally Markov with respect to the given ancestral graph $G$. The *Gaussian ancestral graph model* based on $G$ is defined to be the family of all normal distributions

$$\mathbf{N}(G) = \big(\mathcal{N}_V(0, \Sigma) \mid \Sigma \in \mathbf{P}(G)\big). \quad (1)$$

As shown in RS (2002, §8.4), the model $\mathbf{N}(G)$ forms a curved exponential family.

### 4.1  PARAMETERIZATION

RS (2002, §8) provide a parameterization of the Gaussian ancestral graph model $\mathbf{N}(G)$. This parameterization associates one parameter with each vertex in $V$ and each edge in $E$. Let $\Lambda = (\lambda_{ij})$ be a positive definite $\text{un}_G \times \text{un}_G$ matrix such that $\lambda_{ij} \neq 0$ only if $i = j$ or $i - j$. Recall that $i, j \in \text{un}_G$ can only be adjacent by an undirected edge. Let $\Omega = (\omega_{ij})$ be a positive definite $(V \setminus \text{un}_G) \times (V \setminus \text{un}_G)$ matrix such that $\omega_{ij} \neq 0$ only if $i = j$ or $i \leftrightarrow j$. Finally, let $B = (\beta_{ij})$ be a $V \times V$ matrix such that $\beta_{ij} \neq 0$ only if $j \to i$. Note that the $\text{un}_G \times V$ submatrix of $B$ must be zero, i.e. $B_{\text{un}_G, V} = 0$, because no vertex in $\text{un}_G$ has a directed edge pointing towards it. With the parameter matrices $\Lambda, B, \Omega$, we can define the covariance matrix

$$\Sigma = (I_V - B)^{-1} \begin{pmatrix} \Lambda^{-1} & 0 \\ 0 & \Omega \end{pmatrix} (I_V - B)^{-\mathrm{T}}, \quad (2)$$

which satisfies $\Sigma \in \mathbf{P}(G)$. Equivalently said, the normal distribution $\mathcal{N}_V(0, \Sigma)$ is globally Markov with respect to the considered ancestral graph $G$. If $G$ is maximal, then for any $\Sigma \in \mathbf{P}(G)$ there exist unique $\Lambda, \Omega, B$ of the above type such that (2) holds.

The population interpretation of the parameters is the following: First, the parameter matrix $\Lambda$ clearly forms an inverse covariance matrix for the undirected subgraph $G_{\text{un}_G}$. Second, just as for Gaussian DAG models, the parameter $\beta_{ij}$, associated with a directed edge $j \to i$, is the regression coefficient for variable $j$ in the regression of variable $i$ on its parents $\text{pa}(i)$. Third, the parameter $\omega_{ii}$ is the conditional variance of the conditional distribution $(Y_i \mid Y_{\text{pa}(i)})$, or equivalently $\omega_{ii} = \text{Var}[\varepsilon_i]$, where

$$\varepsilon_i = Y_i - \sum_{j \in \text{pa}(i)} \beta_{ij} Y_j. \quad (3)$$

The parameter $\omega_{ij}$ for $i \leftrightarrow j$ is the covariance between the residuals $\varepsilon_i$ and $\varepsilon_j$. We illustrate the parameterization by showing in Figure 3 the parameters for the graph from Figure 2.

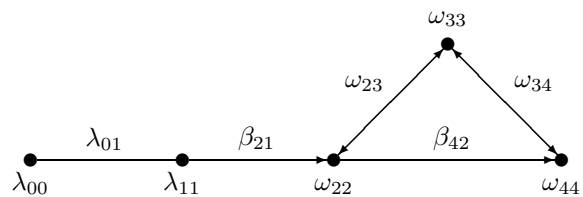

Figure 3: Parameters of a Gaussian ancestral graph model.

### 4.2  MAXIMUM LIKELIHOOD ESTIMATION

This article considers the estimation of the unknown parameter $\Sigma$, or equivalently $\Lambda, \Omega, B$, of a Gaussian ancestral graph model $\mathbf{N}(G)$ based on a sample of i.i.d. observations from $\mathbf{N}(G)$. We assume that $G$ is maximal. Let the i.i.d. copies in the sample be indexed by the set $N$, which can be interpreted as indexing the subjects on which we observed the variables in $V$. Then we can group the observed random vectors in the sample as columns in the $V \times N$ matrix $Y$, which means that $Y_{im}$ represents the observation of the $i$-th variable on the $m$-th subject. The sample size is $n = |N|$ and the number of variables is $p = |V|$.

Since our model assumes a zero mean, the empirical covariance matrix is defined to be

$$S = \frac{1}{n} Y Y^{\mathrm{T}} \in \mathbb{R}^{V \times V}. \quad (4)$$

We shall assume that $n \geq p$ such that $S$ is positive definite with probability one. The case where the model also includes an unknown mean vector $\mu \in \mathbb{R}^V$ can be treated by estimating $\mu$ by the empirical mean vector $\bar{Y} \in \mathbb{R}^V$, i.e. the vector of row means of $Y$. The empirical covariance matrix would then be the matrix

$$\tilde{S} = \frac{1}{n}(Y - \bar{Y} \otimes 1_N)(Y - \bar{Y} \otimes 1_N)^{\mathrm{T}} \in \mathbb{R}^{V \times V}, \quad (5)$$

where $1_N = (1, \ldots, 1) \in \mathbb{R}^N$ and $\otimes$ is the Kronecker product. Estimation of $\Sigma$ would proceed in the same way as in the centered case by using $\tilde{S}$ instead of $S$; the only change being that $n \geq p+1$ ensures almost sure positive definiteness of $\tilde{S}$.

In the centered case considered throughout the remainder of this paper, the density function of $Y$ with respect to the Lebesgue measure is the function $f_\Sigma : \mathbb{R}^{V \times N} \to \mathbb{R}$, which can be expressed as

$$\begin{aligned} f_\Sigma(y) &= (2\pi)^{-np/2} |\Sigma|^{-n/2} \exp\{-\frac{1}{2}\mathrm{tr}(\Sigma^{-1} yy^{\mathrm{T}})\} \\ &= (2\pi)^{-np/2} |\Sigma|^{-n/2} \exp\{-\frac{n}{2}\mathrm{tr}(\Sigma^{-1} S)\}; \end{aligned} \quad (6)$$

see e.g. Edwards (2000, §3.1). Considered as a function of the unknown parameters for fixed data $y$ this gives the likelihood function of the Gaussian ancestral graph model $\mathbf{N}(G)$ as the mapping $L : \mathbf{P}(G) \to \mathbb{R}$ where $L(\Sigma) = f_\Sigma(y)$. In ML estimation, the parameter $\Sigma$ is estimated by the maximizer $\hat{\Sigma}$ of the likelihood $L$. Usually, one considers more conveniently the maximization of the log-likelihood $\ell = \log L$, which, ignoring an additive constant, takes the form

$$\ell(\Sigma) = -\frac{n}{2}\log|\Sigma| - \frac{n}{2}\mathrm{tr}(\Sigma^{-1} S). \quad (7)$$

Positive definiteness of $S$ ensures existence of the global maximum of $\ell(\Sigma)$ over $\mathbf{P}(G)$ but there may exist multiple local maxima (Drton and Richardson 2004).

### 4.3 EMPLOYING THE DECOMPOSITION OF AN ANCESTRAL GRAPH

The decomposition of an ancestral graph $G$ into an undirected and a directed-bidirected part is accompanied by a factorization of the density function of a distribution in the associated model $\mathbf{N}(G)$; cf. RS (2002, §8.5). In detail, if $\Sigma \in \mathbf{P}(G)$, then

$$f_\Sigma(y) = f_\Lambda(y_{\mathrm{un}_G}) f_{B,\Omega}(y_{V \setminus \mathrm{un}_G} \mid y_{\mathrm{un}_G}), \quad (8)$$

where $f_\Lambda(y_{\mathrm{un}_G})$ is the marginal density of $Y_{\mathrm{un}_G}$, and $f_{B,\Omega}(y_{V \setminus \mathrm{un}_G} \mid y_{\mathrm{un}_G})$ is the conditional density of $Y_{V \setminus \mathrm{un}_G}$ given $Y_{\mathrm{un}_G}$. Since the parameters $\Lambda$, $B$ and $\Omega$ are variation-independent, we can find the MLE of $\Lambda$ by maximizing the marginal likelihood function $L(\Lambda) = f_\Lambda(y_{\mathrm{un}_G})$. Since $Y_{\mathrm{un}_G} \sim \mathcal{N}(0, \Lambda)$, this is precisely fitting an undirected graph model based on the graph $G_{\mathrm{un}_G}$ using only the observations for variables in $\mathrm{un}_G$, that is $Y_{\mathrm{un}_G, N}$. Thus, the MLE $\hat{\Lambda}$ of $\Lambda$ can be obtained by iterative proportional fitting (e.g. Whittaker, 1990, pp. 182–185).

In order to find the MLE $(\hat{B}, \hat{\Omega})$ of $(B, \Omega)$ we can maximize the conditional likelihood function

$$L(B, \Omega) = f_{B,\Omega}(y_{V \setminus \mathrm{un}_G} \mid y_{\mathrm{un}_G}). \quad (9)$$

It is easy to see that the global Markov property for the graph $G$ implies that

$$L(B, \Omega) = f_{B,\Omega}(y_{V \setminus \mathrm{un}_G} \mid y_{\mathrm{pa}(V \setminus \mathrm{un}_G) \cap \mathrm{un}_G}). \quad (10)$$

Thus, only the variables in $\mathrm{db}_G = [V \setminus \mathrm{un}_G] \cup \mathrm{pa}(V \setminus \mathrm{un}_G)$ are needed for estimating $(B, \Omega)$; i.e. we need only the observations $Y_{\mathrm{db}_G, N}$. The set $\mathrm{db}_G$ comprises all vertices $i$ in $G$ that are the endpoint of at least one directed or bidirected edge, i.e. $i \to j$, $i \leftarrow j$ or $i \leftrightarrow j$. For $G$ from Figure 2, the induced subgraph $G_{\mathrm{db}_G}$ is shown in Figure 4.

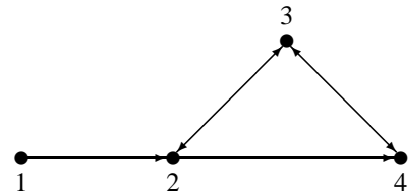

Figure 4: The graph $G_{\mathrm{db}_G}$.

Next, we present an algorithm for estimating $(B, \Omega)$ extending an idea developed in Drton and Richardson (2003), who in fact consider ancestral graphs with bidirected edges only. This idea is to iteratively fit a conditional distribution for a fixed marginal distribution. Thus we call this algorithm *iterative conditional fitting* (ICF). ICF is in nice duality to iterative proportional fitting, in which, cycling through a list of cliques, the marginal distribution over a clique $C \subseteq V$ is fitted for fixed conditional distribution $(Y_{V \setminus C} \mid Y_C)$.

## 5 ITERATIVE CONDITIONAL FITTING

### 5.1 THE GENERAL ALGORITHM

Let $G$ be a maximal ancestral graph. The idea of iterative conditional fitting (ICF) put forward by Drton and Richardson (2003) is to repeatedly iterate through all vertices $i \in V$, and

(i) Fix the marginal distribution for $Y_{-i} = Y_{V \setminus \{i\}}$.

(ii) Fit the conditional distribution $(Y_i \mid Y_{-i})$ under the constraints implied by the Gaussian ancestral graph model $\mathbf{N}(G)$.

(iii) Find a new estimate of $\Sigma$ from the estimated parameters of the conditional distribution $(Y_i \mid Y_{-i})$ and the fixed parameters of the marginal distribution of $Y_{-i}$.

In Drton and Richardson (2003), where the graph $G$ only contained bidirected edges, the problem of fitting $(Y_i \mid Y_{-i})$ under constraints could be rephrased as a least squares regression problem. Here, however, where $G$ also contains directed edges, the consideration of $(Y_i \mid Y_{-i})$ is complicated. Fortunately, we can "remove" directed edges by forming residuals as in (3), and considering the conditional distribution $(Y_i \mid \varepsilon_{-i})$ as presented in the following. Such a residual trick can already be found in Telser (1964).

Before formulating ICF for ancestral graphs, we note two facts. First, the maximization of $L(B, \Omega)$ from (9) is by (8) equivalent to maximizing $L(\Sigma) = L(\Lambda, B, \Omega)$ while holding $\Lambda$ fixed to some feasible value $\Lambda_0$, which could be for example the identity matrix or the matrix found by iterative proportional fitting as described in 4.3. Second, fixing $\Lambda = \Lambda_0$, the matrix $\varepsilon = (I_V - B)Y$ has $i$-th row equal to the residual $\varepsilon_i$ defined in (3) and each column of $\varepsilon$ has covariance matrix $\Psi = (\psi_{ij})$ equal to

$$\Psi = \begin{pmatrix} \Lambda_0^{-1} & 0 \\ 0 & \Omega \end{pmatrix}. \tag{11}$$

From this and the fact $B_{\mathrm{un}_G, V} = 0$ we see that, in order to estimate $(B, \Omega)$ we need only cycle through the vertices $i \notin \mathrm{un}_G$.

Next we compute the conditional distribution $(Y_i \mid \varepsilon_{-i})$ for $i \notin \mathrm{un}_G$. This distribution is obviously Gaussian and its conditional variance equals

$$\mathrm{Var}[Y_i \mid \varepsilon_{-i}] = \omega_{ii.-i}, \tag{12}$$

which is defined as

$$\omega_{ii.-i} = \omega_{ii} - \Omega_{i,-i}(\Omega_{-i,-i})^{-1}\Omega_{-i,i}. \tag{13}$$

Equation (12) holds because, for $i \notin \mathrm{un}_G$

$$\begin{aligned}\mathrm{Var}[Y_i \mid \varepsilon_{-i}] &= \mathrm{Var}[\varepsilon_i \mid \varepsilon_{-i}] \\ &= \psi_{ii} - \Psi_{i,-i}(\Psi_{-i,-i})^{-1}\Psi_{-i,i} \\ &= \omega_{ii} - \Omega_{i,-i}(\Omega_{-i,-i})^{-1}\Omega_{-i,i}.\end{aligned} \tag{14}$$

Note that when writing $\Omega_{-i,-i}$ we mean the $[V \setminus (\mathrm{un}_G \cup \{i\})] \times [V \setminus (\mathrm{un}_G \cup \{i\})]$ submatrix of $\Omega$. The normal distribution $(Y_i \mid \varepsilon_{-i})$ is now specified by (12) and the conditional expectation, which equals

$$\begin{aligned}\mathrm{E}[Y_i \mid \varepsilon_{-i}] &= \sum_{j \in \mathrm{pa}(i)} \beta_{ij}\mathrm{E}[Y_j \mid \varepsilon_{-i}] + \mathrm{E}[\varepsilon_i \mid \varepsilon_{-i}] \\ &= \sum_{j \in \mathrm{pa}(i)} \beta_{ij}Y_j + \sum_{k \in \mathrm{sp}(i)} \omega_{ik}Z_k,\end{aligned} \tag{15}$$

where the *pseudo-variable* $Z_k$ is equal to the $k$-th row in

$$Z_{\mathrm{sp}(i)} = [(\Omega_{-i,-i})^{-1}]_{\mathrm{sp}(i),-i}\,\varepsilon_{-i}. \tag{16}$$

The derivation of (15) relies on two facts. First, for any $j \in \mathrm{pa}(i)$, $Y_j$ is a function of $\varepsilon_{\mathrm{an}(i) \setminus \{i\}}$ and thus, $\mathrm{E}[Y_j \mid \varepsilon_{-i}] = Y_j$. Second, for $i \notin \mathrm{un}_G$ the residual covariance $\psi_{ij} = 0$ if $i \not\leftrightarrow j$. Thus, $\Psi_{i,\mathrm{nsp}(i)} = 0$, which implies that for $i \notin \mathrm{un}_G$:

$$\begin{aligned}\mathrm{E}[\varepsilon_i \mid \varepsilon_{-i}] &= \Psi_{i,-i}(\Psi_{-i,-i})^{-1}\varepsilon_{-i} \\ &= \Psi_{i,\mathrm{sp}(i)}[(\Psi_{-i,-i})^{-1}]_{\mathrm{sp}(i),-i}\,\varepsilon_{-i} \\ &= \Omega_{i,\mathrm{sp}(i)}[(\Omega_{-i,-i})^{-1}]_{\mathrm{sp}(i),-i}\,\varepsilon_{-i} \\ &= \sum_{k \in \mathrm{sp}(i)} \omega_{ik}Z_k.\end{aligned} \tag{17}$$

Now we are ready to formulate ICF for ancestral graphs. Until convergence, for each $i \notin \mathrm{un}_G$:

1. Fix $\Omega_{-i,-i}$ and all $B_{j,\mathrm{pa}(j)} = (\beta_{j\ell} \mid \ell \in \mathrm{pa}(j))$ for $j \neq i$;

2. Use the fixed $\beta_{j\ell}$ to compute the residuals $\varepsilon_j$ for $j \neq i$ from (3);

3. Use the fixed $\Omega_{-i,-i}$ to compute the pseudo-variables $Z_k$ for $k \in \mathrm{sp}(i)$;

4. Carry out a least squares regression with response variable $Y_i$ and covariates $Y_j$, $j \in \mathrm{pa}(i)$, and $Z_k$, $k \in \mathrm{sp}(i)$ to obtain estimates of $\beta_{ij}$, $j \in \mathrm{pa}(i)$, $\omega_{ik}$, $k \in \mathrm{sp}(i)$, and $\omega_{ii.-i}$;

5. Compute an estimate of $\omega_{ii}$ using the new estimates and the fixed parameters from the relation $\omega_{ii} = \omega_{ii.-i} + \Omega_{i,-i}(\Omega_{-i,-i})^{-1}\Omega_{-i,i}$, compare (13).

After steps (1)-(5), we move on to the next vertex in $V \setminus \mathrm{un}_G$, continuing until convergence.

### 5.2 CONVERGENCE

The ICF algorithm is an iterative partial maximization algorithm (Lauritzen 1996, App. A.4) since in the $i$-th step we maximize the conditional likelihood $L(B, \Omega)$ from (9) over the section in the parameter space defined by fixing the parameters $\Omega_{-i,-i}$, and $B_{j,\mathrm{pa}(j)}$, $j \neq i$. It follows that for any feasible starting value the algorithm produces a sequence of estimates whose accumulation points are local maxima or saddle points of the likelihood. Further, evaluating the likelihood at different accumulation points of such a sequence gives the same value.

### 5.3 APPLYING ICF TO DAGS

It is well known that the MLE of the parameters of a Gaussian DAG model can be found by carrying out a finite number of regressions (e.g. Goldberger 1964, or Andersson and Perlman 1998). DAGs are special cases of ancestral graphs so we can also apply ICF to a Gaussian DAG model. If the graph $G$ is a DAG then $\mathrm{sp}(i) = \emptyset$ for all $i \in V$.

Therefore, the conditional distribution of $(Y_i \mid \varepsilon_{-i})$ is fitted by regressing solely on the parents $Y_j$, $j \in \text{pa}(i)$; cf. (15). Thus the least squares regression carried out in the $i$-th step of ICF is always the same since it involves no pseudo-variables, which could change from one iteration to the other. This means that ICF reduces to the standard approach of fitting Gaussian DAG models.

### 5.4 ICF IN AN EXAMPLE

We illustrate estimation of $(B, \Omega)$ by ICF in the example of the ancestral graph depicted in Figure 2. The set $\text{un}_G$ equals $\{0, 1\}$ and in fact only the variables $\text{db}_G = \{1, 2, 3, 4\}$ are relevant for estimating $(B, \Omega)$, compare Figure 4. The iteration steps, described in items (1)-(5) in §5.1, have to be carried out only for $i \in V \setminus \text{un}_G = \{2, 3, 4\}$. In Figure 5, we show the response variable $Y_i$ in the $i$-th ICF update step as a filled circle. The remaining variables are depicted as unfilled circles. A vertex labelled by a number $j$ represents the random variable $Y_j$ and a vertex labelled by $Z_j$ represents the pseudo-variable defined in (16). Finally, we use directed edges pointing from a covariate $Y_j$ or $Z_j$ to the response $Y_i$ to indicate the structure of the least squares regression that has to be performed in the $i$-th step of ICF. Note that we suppress the irrelevant variable 0.

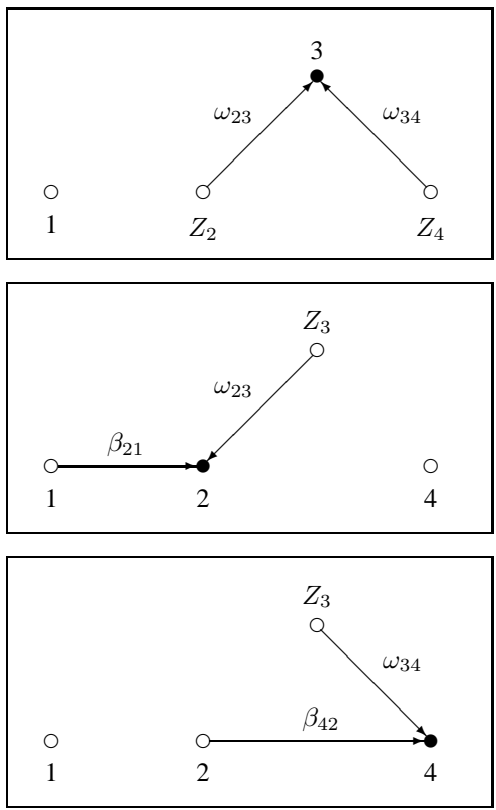

Figure 5: Illustration of the ICF update steps.

## 6 AN IMPLEMENTATION

The statistical programming language R (Ihaka and Gentleman 1996) provides a freeware environment for programming in interpreted code building on a large number of available routines. As part of the "graphical models in R" initiative (Lauritzen 2002), Marchetti and Drton developed a function library called 'ggm', which implements functions for fitting Gaussian graphical models and, in particular, provides an implementation of ICF. The package can be downloaded from http://cran.r-project.org/.

Here we show an example session in R using 'ggm' and data on noctuid moth trappings from the R library 'SIN', which implements the model selection method described in Drton and Perlman (2004). The data comprise $n = 72$ measurements for each one of six variables: the (log-transformed) number of moths caught in a light trap in one night (moth), the minimum night temperature (min), the previous day's maximum temperature (max), the average wind speed during the night (wind), the amount of rain during the night (rain), and the percentage of starlight obscured by clouds (cloud); compare Whittaker (1990, §10.3). We begin by loading 'ggm', 'SIN', and the data, for which we display the correlation matrix and the sample size.

```
> library(ggm); library(SIN)
> data(moth)
> moth$corr
        min   max  wind  rain cloud  moth
min    1.00  0.40  0.37  0.18 -0.46  0.29
max    0.40  1.00  0.02 -0.09  0.02  0.22
wind   0.37  0.02  1.00  0.05 -0.13 -0.24
rain   0.18 -0.09  0.05  1.00 -0.47  0.11
cloud -0.46  0.02 -0.13 -0.47  1.00 -0.37
moth   0.29  0.22 -0.24  0.11 -0.37  1.00
> moth$n
[1] 72
```

Concentrating on the five variables max, wind, rain, cloud, and moth we fit the model induced by the (maximal) ancestral graph from Figure 2. We first define the graph by specifying complete subsets for the undirected and bidirected subgraphs, and regression structures for the directed subgraph. The resulting graph is represented by an adjacency matrix $A = (a_{ij})$ where $a_{ij} = a_{ji} = 1$ if $i - j$, $a_{ij} = a_{ji} = 2$ if $i \leftrightarrow j$, and $a_{ij} = 1$, $a_{ji} = 0$ if $i \to j$.

```
> mag <- makeAG(ug=UG(~wind*rain),
+          dag=DAG(cloud~rain, moth~cloud),
+          bg=UG(~max*cloud+max*moth))
> mag
      max cloud moth wind rain
max     0     2    2    0    0
cloud   2     0    1    0    0
moth    2     0    0    0    0
wind    0     0    0    0    1
rain    0     1    0    1    0
```

The package 'ggm' also provides a rudimentary tool for plotting graphs.

```
> drawGraph(mag)
```

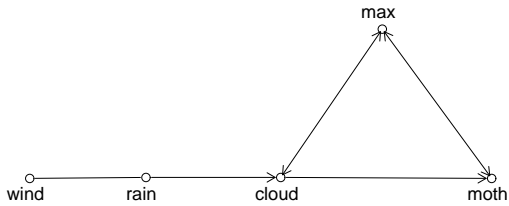

Now we are able to fit the Gaussian ancestral graph model to the data using the function 'fitAncestralGraph', which returns $\hat{\Sigma}$ as Shat, $\hat{\Lambda}$ as Lhat, $I_V - \hat{B}$ as Bhat and $\hat{\Omega}$ as Ohat. The output also includes the deviance statistic dev, the degrees of freedom df and the number of iterations it, that is, the number of full cycles through all $i \notin \text{un}_G$ during ICF. The deviance statistic is computed as

$$\text{dev} = 2\ell(S) - 2\ell(\hat{\Sigma}). \qquad (18)$$

Note, however, that the ICF estimate $\hat{\Sigma}$ may only be a local maximum (Drton and Richardson 2004).

```
> icf <- fitAncestralGraph(mag, moth$corr,
+                          moth$n)
> lapply( icf , round, 2 )
$Shat
         max  wind  rain cloud  moth
  max   1.00  0.00  0.00 -0.02  0.23
  wind  0.00  1.00  0.05 -0.02  0.01
  rain  0.00  0.05  1.00 -0.47  0.18
  cloud -0.02 -0.02 -0.47  1.00 -0.38
  moth  0.23  0.01  0.18 -0.38  1.01

$Lhat
      max wind rain cloud moth
max     0 0.00 0.00     0    0
wind    0 1.00 0.05     0    0
rain    0 0.05 1.00     0    0
cloud   0 0.00 0.00     0    0
moth    0 0.00 0.00     0    0

$Bhat
      max wind rain cloud moth
max     1    0 0.00  0.00    0
wind    0    1 0.00  0.00    0
rain    0    0 1.00  0.00    0
cloud   0    0 0.47  1.00    0
moth    0    0 0.00  0.38    1

$Ohat
       max wind rain cloud moth
max   1.00    0    0 -0.02 0.23
wind  0.00    0    0  0.00 0.00
rain  0.00    0    0  0.00 0.00
cloud -0.02   0    0  0.78 0.00
moth  0.23    0    0  0.00 0.86

$dev
[1] 10.22
$df
[1] 5
$it
[1] 6
```

Comparing the deviance and the degrees of freedom using the asymptotic distribution of the deviance as $\chi^2_{\text{df}}$ yields a p-value of $0.07$ suggesting that the model is not inappropriate. However, the inclusion of the additional directed edge wind $\to$ moth greatly improves the fit with a deviance of $2.01$ over $4$ degrees of freedom and an associated p-value of $0.73$.

## 7   EXPERIMENTS

We illustrate the performance of ICF using simulated data from a chordless bidirected cycle as shown in Figure 6. In detail, we chose to simulate data from a multivariate normal distribution $\mathcal{N}_V(0, \Sigma_V)$ with covariance matrix

$$\Sigma_V = \begin{pmatrix} 1 & 0.3 & & & 0.3 \\ 0.3 & 1 & 0.3 & & \\ & \ddots & \ddots & \ddots & \\ & & 0.3 & 1 & 0.3 \\ 0.3 & & & 0.3 & 1 \end{pmatrix} \in \mathbb{R}^{V \times V}, \qquad (19)$$

where only non-zero entries are shown. We varied the number of variables $p = |V|$ from $p = 10$ to $p = 100$ and for each $p$, we simulated 100 samples of size $n = p + 30$. For each one of the samples we ran ICF stopping the algorithm if, from one iteration to the next, the maximum absolute deviation between any entry in the estimated covariance matrix was smaller than $10^{-6}$. Surprisingly, the mean number of iterations in ICF, i.e. full cycles through $V$, remained stable for the different $p$, varying only in the interval $[7.1, 7.5]$. However, each iteration consists of $p$ updates steps and becomes obviously more complex for increasing $p$. Figure 7 shows how the usage of CPU time on a 2.4 GHz Pentium IV PC increases with growing $p$.

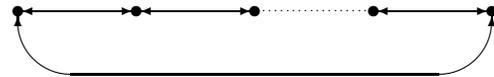

Figure 6: Chordless bidirected cycle.

## 8   CONCLUSION

Iterative conditional fitting (ICF) is an iterative partial maximization algorithm for fitting Gaussian ancestral graph models. Fitting conditional distributions while fixing marginal distributions, ICF is in duality to iterative proportional fitting, in which marginal distributions are fitted while conditional distributions are fixed. ICF is particularly attractive since if the ancestral graph under consideration is in fact a DAG, then the likelihood is maximized in a finite number of steps performing exactly the regressions commonly used for fitting Gaussian DAG models.

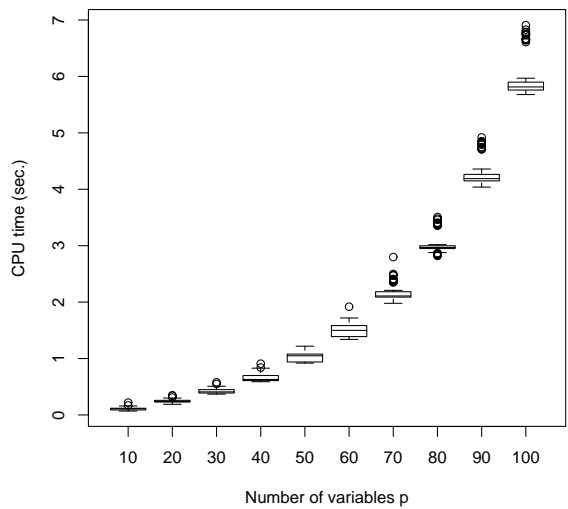

Figure 7: Usage of CPU time in ICF for growing number of variables $p$.

The ICF algorithm has similarities with the Iterative Conditional Modes (ICM) algorithm of Besag (1986). However, ICM obtains maximum *a posteriori* estimates in a Bayesian framework, whereas our ICF maximizes a likelihood function. The difference in the update steps is that in ICM conditional density functions are maximized, whereas in ICF one maximizes conditional likelihood functions.

Another related algorithm is the Conditional Iterative Proportional Fitting (CIPF) algorithm of Cramer (1998, 2000). CIPF can be used to fit a model that comprises joint distributions for which a set of conditional distributions are set equal to prescribed conditionals. However, CIPF differs from ICF because the update steps of ICF do not simply equate a conditional distribution with a prescribed conditional, but rather find a conditional distribution by maximizing a conditional likelihood function. In particular, these maximizers will generally not be the same in two different iterations of ICF.

A topic of future work will be using Markov equivalence of ancestral graphs for improving the efficiency of ICF. As it is true for DAGs, different ancestral graphs may induce the same statistical model, in which case the graphs are called Markov equivalent. Since the update steps of the ICF algorithm depend on the graph itself, it is important to determine which graph in a whole class of Markov equivalent graphs allows for the most efficient fitting of the associated model (see also Drton and Richardson 2003, §4.2.4).

Finally, ICF has the nice feature that its main idea of decomposing the complicated overall maximization problem into a sequence of simpler optimization problems seems also promising for the development of fitting methodology in the case of discrete variables.

**Acknowledgements**

We thank Steffen Lauritzen for pointing out the duality between ICF and IPF and acknowledge support by NSF grants DMS 9972008 and DMS 0071818.